\documentclass[longbibliography,prd,nofootinbib,amsfonts,notitlepage]{revtex4-2}
\usepackage[utf8]{inputenc}
\usepackage[T1]{fontenc}
\usepackage{hyperref}
\usepackage{graphicx,bm,amsmath,color}
\usepackage{bbold}
\usepackage{wasysym}
\usepackage{verbatim}
\usepackage{amssymb}
\usepackage{mathrsfs}
\usepackage{epstopdf}
\usepackage{animate}
\usepackage{xmpmulti}
\usepackage{centernot}
\usepackage{multirow}
\usepackage{tikz}
\usepackage{graphicx}
\usepackage{float}
\usepackage{mathtools}
\usepackage{comment}
\DeclareMathOperator{\arcch}{arccosh}
\DeclareMathOperator{\arcsh}{arcsinh}

\DeclareMathOperator{\sech}{sech}

\makeatletter

\newcommand{\be}{\begin{equation}}
\newcommand{\ee}{\end{equation}}
\newcommand{\beq}{\begin{eqnarray}}
\newcommand{\eeq}{\end{eqnarray}}
\newcommand{\ba}{\begin{align}}
\newcommand{\ea}{\end{align}}

\newcommand{\blue}[1]{{\leavevmode\color{blue}{#1}}}
\newcommand{\mathi}{{\rm i}}
\newcommand{\freq}{\omega_p}
\newcommand{\measure}{\frac{{\rm d}^3p}{\freq}}

\begin{document}

\title{Geometrizing the Klein--Gordon and Dirac equations in Doubly Special Relativity}

\author{S.~A. Franchino-Viñas}
\address{Departamento de F\'isica, Facultad de Ciencias Exactas
Universidad Nacional de La Plata, C.C.\ 67 (1900), La Plata, Argentina}

\address{Institut für Theoretische Physik, Universität Heidelberg, D-69120 Heidelberg, Germany}

\address{Helmholtz-Zentrum Dresden-Rossendorf, Bautzner Landstraße 400, 01328 Dresden, Germany.}
\email{s.franchino-vinas@hzdr.de}

\author{J.J. Relancio}
\affiliation{Departamento de Física, Universidad de Burgos, 09001 Burgos, Spain;\\Departamento de F\'{\i}sica Te\'orica and Centro de Astropartículas y F\'{\i}sica de Altas Energ\'{\i}as (CAPA),
Universidad de Zaragoza, Zaragoza 50009, Spain
}
\email{jjrelancio@ubu.es}

\begin{abstract}
In this work we discuss the deformed relativistic wave equations, namely the Klein--Gordon and Dirac equations in a Doubly Special Relativity scenario. 
We employ what we call a geometric approach, based on the geometry of a curved momentum space, which should be seen as complementary to the more spread algebraic one. 
In this frame we are able to rederive well-known algebraic expressions, as well as to treat yet unresolved issues, to wit, the explicit relation between both equations, the discrete symmetries for Dirac particles, the fate of covariance, and the formal definition of a Hilbert space for the Klein--Gordon case.
\end{abstract}

\maketitle

\section{Introduction}
A quantum gravity theory (QGT), i.e. a theory encompassing quantum field theory (QFT) and general relativity (GR), has been looked for during several decades. 
The (probably) simplest trial, a quantum theory of gravitation where the mediation of the interaction is carried out by the graviton, a spin-2 particle, has lead to several inconveniences~\cite{Goroff:1985th,Goroff:1985sz}, which ended up in its understanding as an effective field theory~\cite{Donoghue:1994dn}. 
In spite of the success of the latter, there remain many subtleties, for example those related to the renormalization process and the gauge-independence of the computations~\cite{Giacchini:2020dhv,Giacchini:2020zrl,Solodukhin:2020vuw}.

Searching for a more fundamental theory and  principles, the community has developed several theoretical frameworks, such as string theory~\cite{Mukhi:2011zz,Aharony:1999ks,Dienes:1996du}, loop quantum gravity~\cite{Sahlmann:2010zf,Dupuis:2012yw}, causal dynamical triangulations \cite{Loll_2019}, causal set theory~\cite{Wallden:2013kka,Wallden:2010sh,Henson:2006kf} and functional renormalization group~\cite{Bonanno:2020bil}.  In (almost) all of them a minimum length arises~\cite{Gross:1987ar,Amati:1988tn, Garay1995,Chang:2011jj},  which is heuristically associated with the Planck length $\ell_P \sim 1.6\times 10^{-33}$\,cm (or Planck mass $M_P \sim 1.22\times 10^{19}$\,GeV); see~\cite{Hossenfelder:2012jw} for a review of minimal length scale scenarios.
 This small length (high-energy scale) is expected to somehow separate the regime where spacetime displays its classical nature from the one where it develops a quantum structure.

One can naturally guess that all of the above-mentioned theories should lead to novel scenarios with possible observable implications. 
However, the description of the latter is generally arduous. In order to enhance the connection with the observational side,
another way of thinking has arisen not long ago: instead of considering a fundamental theory of quantum gravity, one can explore a low-energy limit of it, leading indeed to testable phenomenology~\cite{Addazi:2021xuf}. 

A possible way to follow this bottom-up approach is through a modification of the special relativistic kinematics by introducing a high-energy scale. 
There are two different possibilities of doing this,
depending on which is the fate of Lorentz symmetry: for high energies,
one can consider that it is either violated or deformed. The former scenario is implemented in the framework of Lorentz invariance violation (LIV)~\cite{Colladay:1998fq,Kostelecky:2008ts,Mattingly:2005re,Liberati:2013xla}, while the latter is inherent to Doubly (or Deformed) Special Relativity (DSR) theories~\cite{AmelinoCamelia:2008qg}. An immediate consequence is that the relativity principle characterizing Special Relativity (SR) is lost in LIV, while simply deformed in DSR.    

Both ideas have been extensively developed from  an algebraic approach.
Indeed, the original proposal by Snyder
 consisted in a deformed algebra of position and momentum operators, in an attempt to regularize ultraviolet divergences arising in QFT~\cite{Snyder:1946qz,Snyder:1947nq}.
More recently, after the introduction of the  $q$-deformed Poincaré algebra through a  Drinfeld--Jimbo deformation~\cite{Lukierski:1991pn}, several works have been devoted to understand $\kappa$-Minkowski and Snyder algebras \cite{Kosinski:2001ii,Kowalski-Glikman:2003qjp, Mignemi:2008kn,Govindarajan:2009wt, Mignemi:2011gr,Poulain:2018two,Arzano:2020jro,Lizzi:2021rlb},
including also proposals on how to obtain a Dirac equation~\cite{Lukierski:1992dt, Giller:1992xg, Nowicki:1992if, Agostini:2002yd, DAndrea:2005hjg}.

In this article we will employ a less developed, geometric point of view,
to discuss generalizations of the  Klein--Gordon and Dirac equations,
what may be appreciated by readers as an efficient and intuitive alternative. Recently,  a clear connection between the geometry in momentum space and noncommutative physics was shown to exist~\cite{Carmona:2021gbg,Relancio:2021ahm,Wagner:2021bqz,Wagner:2021thc}. This idea was already latent in the original paper of Born~\cite{Born:1938}, where a first proposal of curved momentum space was made and a lattice structure for spacetime was shown to arise. 
These considerations are englobed in the broader context  
of  velocity or momentum dependent spacetime, the so-called Finsler and Hamilton geometries~\cite{miron2001geometry}. Among others, there are several works in LIV setups describing the propagation of particles with a modified dispersion relation through a Finsler spacetime~\cite{Kostelecky:2011qz,Barcelo:2001cp,Weinfurtner:2006wt,Hasse:2019zqi,Stavrinos:2016xyg}. Also in DSR scenarios there exist contributions which consider a velocity~\cite{Girelli:2006fw,Amelino-Camelia:2014rga,Letizia:2016lew} and momentum~\cite{Barcaroli:2016yrl,Barcaroli:2015xda,Barcaroli:2017gvg} dependent geometries, taking into account a deformed dispersion relation.   



Our construction below is greatly motivated by the results in~\cite{Carmona:2019fwf}, where it was shown that all the ingredients of a relativistic deformed kinematics (both a deformed dispersion relation and a composition law) can be singled out in the case of a maximally symmetric (curved) momentum space. In particular,  $\kappa$-Poincaré kinematics\footnote{When in the following we mention $\kappa$-Poincaré we mean only the kinematical momentum sector associated to the deformation, i.e., we do not consider the noncommutative  spacetime coordinates appearing in the construction of such model but only the deformed composition law, the dispersion relation and Lorentz transformations.} can be obtained from a de Sitter space by identifying the deformed composition law, the deformed Lorentz transformations and the deformed dispersion relation: they correspond respectively to the translation isometries,  the Lorentz isometries  and the squared distance to the origin. 
Notice that the last two facts have been  previously contemplated in Refs.~\cite{AmelinoCamelia:2011bm,Gubitosi:2011hgc, Lobo:2016blj}, while the connection between    DSR and a curved momentum space was previously realized in \cite{Kowalski-Glikman:2002oyi,Magueijo:2002xx}.  Also, it is important to keep in mind that different bases of $\kappa$-Poincaré~\cite{Kowalski-Glikman:2002oyi} can be obtained from different choices of momentum coordinates in a de Sitter space~\cite{Carmona:2019fwf}.   

The main difference of~\cite{Carmona:2019fwf} with respect to previous works in that one is able to describe within the same geometrical framework all the ingredients of a relativistic deformed kinematics, viz. a deformed dispersion relation, a deformed composition law, and some Lorentz transformations in the one- and two-particle system. 
Moreover, this construction allows one to describe the more studied kinematics beyond  $\kappa$-Poincaré, such us Snyder kinematics~\cite{Battisti:2010sr} and the so-called hybrid models~\cite{Meljanac:2009ej}. Also, the construction of~\cite{Carmona:2019fwf} can be  generalized in order to include a curvature in spacetime while preserving the deformed kinematics~\cite{Relancio:2020zok,Relancio:2020rys,Pfeifer:2021tas}. 

In order to discuss the modified deformed relativistic wave equations in a geometrical framework, we organize our  article as follows. In Sec.~\ref{sc:KG} we will show how one can introduce the Klein--Gordon equation in curved momentum space. Then, in Sec.~\ref{sec:Dirac} we generalize the technique to a fermionic system of spin one-half. In both cases we show that we reproduce  firmly set results previously obtained within the scheme of Hopf algebras;
additionally, we show that one can implement a variational principle in momentum space by defining appropriate actions. Moreover, we are also able to describe the Dirac equation for Snyder kinematics, which is not contained within the Hopf algebra structure.
We discuss the discrete fermionic symmetries in Sec.~\ref{sec:discrete}. In Sec.~\ref{sec:hilbert}, we consider some formal aspects regarding the definition of the corresponding Hilbert space.
Finally, we state our conclusions in 
Sec.~\ref{sec:conclusions}.

We use the following conventions. We define the Minkowski metric ($\eta_{\mu\nu}$) with mostly minus signs; all other metrics will possess the same signature. Greek indices  are used to label  spacetime components of a tensor ($\mu,\,\nu,\,\cdots=0,1,2,3$), while latin indices  denote just spatial components ($i,\,j,\,\cdots=1,2,3$). The first latin characters ($a,\,b,\,\cdots=0,1,2,3$) are employed for components in the local orthonormal frame given by the (inverse) vierbein $e_{\mu}{}^a$. Regarding momenta, i.e. coordinates, we use the following notation: we denote  $p^2:=p_\mu \eta^{\mu\nu}p_\nu$; the set of all spatial components of a vector $p$ is written as $\vec{p}$ and $\vec{p}^2:= \vec{p}\cdot \vec{p}$. We use units  $\hbar=c= 1$.

\section{Klein--Gordon equation}\label{sc:KG}
The fact that a nontrivial momentum metric  is able to describe the kinematics of DSR has been thoroughly discussed in~\cite{Carmona:2019fwf}. 
Following those lines, in this section we propose a generalization of the usual   Klein--Gordon (KG) equation  to the case in which the metric is momentum dependent.
Let us first recall some basic aspects of the usual case. 
In SR, the KG operator acts on a wave function $\phi$ as
\begin{equation}
   \left(\eta^{\mu \nu} \frac{\partial}{\partial x^\nu}\frac{\partial}{\partial x^\mu} +m^2\right) \phi(x)\,=\,0\,,
   \label{eq:KG_SR}
\end{equation}
where $m$ is the mass of the particle and $\eta_{\mu\nu}$
is the metric in Minkowski spacetime. 
Since its construction relies in the principle of relativity, this equation is devised as invariant under Poincaré symmetries. In addition it is also invariant under general diffeomorphisms in space, which is clear once one recognizes the Beltrami--Laplace operator.
The most general solution to Eq. \eqref{eq:KG_SR} can be written in terms of plane-wave solutions. Transforming to Fourier space we have
\begin{equation}
   \phi(x)\,=\,\frac{\sqrt{2}}{(2\pi)^{3}}\int  {\rm d}^4 p\, e^{\mathi x^\lambda p_\lambda} 
   \tilde \phi(p)\, \delta(C_{\rm M}(p)-m^2)
   \,,
   \label{eq:KG_field}
\end{equation}
where we have defined  the  usual Casimir,
\begin{equation}\label{eq:casimir_M}
  C_{\rm M}(p)\,:=\,p^2\,=\,p_\mu \eta^{\mu\nu}p_\nu\,,
\end{equation}
 and 
 the Dirac delta function enforces the on-shell condition (or dispersion relation) to be satisfied. Some additional overall factors where chosen to simplify the notation later.
Alternatively, the KG equation can be seen as an algebraic equation in momentum space,
\begin{align}
    (C_{\rm M}(p)-m^2) \tilde \phi (p)=0.
\end{align}

Once we introduce a nontrivial pseudo-Riemannian metric $g_{\mu\nu} (p)$ which depends on the momentum, we are constrained to deform the Casimir in a consistent way. In~\cite{Relancio:2020zok} it was shown that, defining the Casimir as  the pseudo-squared distance  to the origin ($p^*$) in momentum space, which we will denote by $ C_\text{D}(p)$, the following relation between the metric and the Casimir holds 
  \begin{align}
 C_\text{D}(p)\,&=\,f^{\mu} g_{\mu \nu }(p) f^{\nu}\,
\label{eq:casimir_metric},
\\
f^{\mu}  (p)\,:&=\,\frac{1}{2} \frac{\partial C_\text{D}(p)}{\partial p_\mu}\,.
   \label{eq:f_definition}
  \end{align}
Notice that our choice for $C_{\rm D}(p)$  can be understood in terms of Synge's world function $\sigma(p',p)$~\cite{Synge:1960ueh}, which is a biscalar equal to one half of the square of the (geodesic) distance between $p'$ and $p$. 
Under this interpretation, we are fixing the first argument of $\sigma$ to be $p^*$.
Moreover, 
the  functions $f^{\mu}$ are the covariant derivatives of Synge's function, $\sigma^{,\mu}$, and will be called generalized momenta because of their role in generating the  Casimir in Eq.~\eqref{eq:casimir_metric} (see also the following sections).
Note thus that this discussion is valid for any  reasonable pseudo-Riemannian metric, as has been shown for example in~\cite{Synge:1960ueh, DeWitt:1965}. 

In SR, i.e. considering a flat metric, expression \eqref{eq:casimir_metric} can be trivially shown to be satisfied by the Casimir defined in Eq. \eqref{eq:casimir_M}.
Instead, in DSR, when regarding the Casimir as the squared of the distance in momentum space,  $f^\mu$ will be nontrivial functions of the momenta. 
This is the case also in the so-called classical basis of $\kappa$-Poincaré: even if in the algebraic context it is considered that the Casimir in this basis is the one of SR~\cite{Borowiec2010}, one can easily see that the momentum metric describing this kinematics leads  to a nontrivial (squared) distance in momentum space~\cite{Pfeifer:2021tas}. 

In the classical  basis of $\kappa$-Poincaré, the distance $C_{\rm D}(p)$ turns out to be a function of $C_{\rm M}$. In an Euclidean setup, where both are positive, one can invert the relation to show that $C_{\rm M}$ may also play the role of a Casimir.
In the Minkowskian case, one can appeal to a Wick rotation to establish the same conclusion.
These different Casimirs then become equivalent (at least at the classical level) to a joint redefinition of all the particles' masses. However, by regarding the Casimir as a propagator in QFT, we expect a different behaviour of the theory in the ultraviolet (UV) regime, i.e. for large momenta. 
One important argument that favours our choice ($C_{\rm D}$) is that, as we will see, then the Klein--Gordon operator equals the ``squared'' Dirac one, without the need of introducing any extra function of the mass.

We may now proceed in a canonical way,
meaning that the generalized  KG equation should enforce the dispersion relation given by the generalized Casimir in Eq. \eqref{eq:casimir_metric}.
The appropriate definition is thus seen to be\footnote{From now on we drop the tilde over $\phi$, since the basic wave function will be defined in momentum space. } 
\begin{equation}
\left(  f^\mu ({p})g_{\mu \nu} ({p})  f^\nu ({p})  -m^2\right) \phi(p)\,=\,0\,.
   \label{eq:KG_DSR}
\end{equation}
The case of SR is simply recovered by considering a flat metric and  the undeformed Casimir $C_{\rm M}$.  

More formally, from a quantum mechanical perspective, expression \eqref{eq:KG_DSR}
should be thought as a representation of the KG equation in momentum space. This entails considering a construction of momentum eigenstates in curved momentum space, a vision dual to  DeWitt idea's of nonrelativistic quantum mechanics in curved configuration space \cite{DeWitt:1952js, DeWitt:1957at}. 
The fact that we want to consider the relativistic case is by far nontrivial; we will offer a more detailed discussion of these issues in Sec. \ref{sec:hilbert}.

By construction Eq. \eqref{eq:KG_DSR} is well-defined, i.e. covariant, under the action of diffeomorphisms in momentum space, since the Casimir has been defined as a distance.
This assertion is true for any metric.
Moreover, Eq. \eqref{eq:KG_DSR}  satisfies a deformed Lorentz invariance if we choose a de Sitter metric as in DSR.
Indeed, in the context of (classical, i.e. not quantum) DSR, there exists a notion, albeit deformed, of Lorentz transformations~\cite{AmelinoCamelia:2008qg}. The action of the latter is such that the Casimir is left invariant and the metric transforms appropriately as a (2,0)-tensor~\cite{Carmona:2019fwf}. 
Once we assume that the field transforms as a scalar under diffeomorphisms $p\to p^\prime$,
i.e.
 \begin{equation}
\phi^\prime (p^\prime)\,=\,\phi(p)\,,
   \end{equation}   
the action of the quantum generators of Lorentz invariance in momentum space can be seen to be given by simple multiplication with their classical expression as given in~\cite{Carmona:2019fwf}.

Contrary to what happens in SR, where the Lorentz generators $J_{\mu\nu}$ are linear functions of $p$, in DSR the Lorentz transformations are generally nonlinear. However, the  equation 
\begin{equation}
\label{eq:casimir_diffeo}
C_{\rm D}(p)\,=\,C_{\rm D}(p')
\end{equation}
holds if $p$ and $p'$ are connected through a deformed Lorentz transformation. This shows that our proposal for the KG equation is invariant under  Lorentz transformations which are the isometries of the metric defining the kinematics of DSR.

\subsection{Klein--Gordon equation in the symmetric basis of  \texorpdfstring{$\kappa$}{k}-Poincaré}
As an example of the above-derived equations, we will write the KG equation in the symmetric basis of $\kappa$-Poincaré  (also called the standard basis~\cite{LUKIERSKI1991331}). Using the formalism developed in \cite{Carmona:2019fwf} one can obtain all the relevant geometric quantities.
 Then, the distance in momentum space can be readily obtained by employing Eq.~\eqref{eq:casimir_metric} in conjunction with\footnote{In a Casimir,  superscripts denote the choice of coordinates, while subscripts refer to the chosen Casimir. For example, the subscript A will label the Casimir considered in the algebraic approach~\cite{Lukierski:1992dt},
 while subscript D refers to the squared distance in momentum space, cf. \eqref{eq:casimir_metric}.}~\eqref{eq:metric_alg}: 
\be
C^{{(S)}}_{\rm D}(p)\,=\, \Lambda^2 \arcch^2 \left(\cosh \left(\frac{p_0}{\Lambda}\right) -\frac{\vec{p}^2}{2 \Lambda^2}\right)\,.
\label{eq:cas_alg_dis}
\ee

One can compare this result with the proposal  in~\cite{Lukierski:1992dt}, which was obtained in an algebraic context. From their definition of Casimir, the associated KG equation is found to be 
\be
\left(  C^{(S)}_\text{A}(p)  -m^2  \right) \phi(p)\,:=\,\left(  \left(2 \Lambda \sinh \left(\frac{p_0}{2\Lambda}\right)\right)^2 - \vec{p}^2  -m^2  \right) \phi(p)\,.
\label{eq:kg_alg}
\ee
Evidently, Eq. \eqref{eq:kg_alg}  differs from our proposal \eqref{eq:cas_alg_dis} of constructing the KG equation from the squared distance in momentum space. However, similar to the situation discussed in the previous subsection, one can  indeed see that the following relation between them exists:
\begin{align}
C^{(S)}_{\text{D}}(p)\,=\,\Lambda^2 \arcch^2\left(1 +\frac{C_{\text{A}}(p)}{2 \Lambda^2}\right)\,.
\end{align}
As commented previously, this difference in the definition of the Casimir should be appreciable in the UV regime after the application of a quantization procedure.  Additionally, as will be seen in Sec. \ref{sec:dirac_symmetric_basis}, our proposal provides a direct link between Dirac and KG equations.

\subsection{An action for scalar fields}\label{sec:action_KG}

Assuming a principle of stationary action, the KG equation~\eqref{eq:KG_DSR} can be obtained from the action
\begin{align}\label{eq:KG_action}
S_{\rm KG}\,:&=\,\int {\rm d}^4p\, \sqrt{-g} \,\phi^*(p) \left(  C_{\rm D}(p)  -m^2  \right)  \phi(p)
\,.
   \end{align}
The factor $\sqrt{-g}$ guarantees invariance under a change of momentum basis. 
This action may serve as a departing point to construct a free quantum field theory,
as has been  proposed in the work by Gol'fand \cite{Golfand:1959vqx, Golfand:1962kjf}, and refined by Mir-Kasimov
\cite{Mir-Kasimov:1966a, Mir-Kasimov:1966b, Mir-Kasimov:1967} and others \cite{Kadyshevsky:1983yc,Donkov:1984fj}.
Indeed, one can first define the theory in a Riemannian momentum space, where the action is positive defined, and afterwards employ an analogue of the Wick-rotation to consider pseudo-Riemannian spaces \cite{Golfand:1962kjf}.

Although this theory would share many features with the proposal of Gol'fand, there is a fundamental difference: we propose the Casimir as a distance in momentum space, whereas in \cite{Golfand:1962kjf} just $p^2$ in the Beltrami chart~\cite{Pascu:2012yu} is employed (which does not correspond to the squared distance in that coordinates).
Notice that, as commented previously, even if we would work with the same coordinates, the models are expected to differ in the UV sector because of the different choice of the Casimir. Moreover, as we will see, our choice of Casimir is the only one allowing to make a direct identification between the KG and Dirac equations\footnote{Actually, the proposal in~\cite{Donkov:1984fj} can be ``squared'' but just to obtain one component of the KG operator in~\cite{Kadyshevsky:1983yc}: the latter proposes a separation of the field into two components, one which vanishes on the mass shell and one that doesn't.}. 

A subtle point arises when one tries to introduce interactions. Many proposals of QFT in $\kappa$-Minkowski and Snyder space have been done in the literature~\cite{Dimitrijevic:2003wv, Freidel:2007hk, Meljanac:2010ps, Girelli:2010wi, Meljanac:2011cs,
Meljanac:2017grw, Meljanac:2017ikx,Mercati:2018hlc, Franchino-Vinas:2018jcs, Arzano:2020jro}, 
most of them relying on the defnition of an appropriate Moyal--Groenewold product (also called star-product) \cite{Groenewold:1946kp, Meljanac:2010et, Mercati:2010qhd, Meljanac:2021qgq}.
The latter encodes the nontrivial addition of momenta and, generally speaking, can be built in a case by case analysis. 
In our case, the adoption of the translation operators \cite{Carmona:2019fwf} seems better suited. We will leave a concrete analysis to a future publication.

\section{Dirac equation}\label{sec:Dirac}
Now that we have illustrated our ideas on how the geometrical approach works in the scalar case, 
a generalization to Dirac particles of spin $1/2$ seems natural. 
Since we are working in curved momentum space, a natural approach is to propose a local  representation of the Lorentz group, $SO(3,1)$.
Our proposal in curved momentum space for the Dirac equation, in momentum space and as an algebraic equation, is thus given by
   \begin{equation}
 \left( \underline{\gamma}^\mu f_\mu({p})-m\right)\psi(p)\,=\,0\,,
 \label{eq:Dirac_DSR}
\end{equation}
where the functions $f_\mu$ are obtained from the covariant functions in Eq. \eqref{eq:f_definition} by a contraction with the inverse metric,
\begin{equation}
 f_\mu(p)\,:=\,g_{\mu\nu}(p)f^\nu(p)\,.
\end{equation}
The curved-momentum-space gamma matrices (with greek indices and underlined) 
are defined in terms of the usual Dirac matrices $\gamma^a$ (with latin indices, with $a=0,1,2,3$), the latter providing a spin-$1/2$ finite-dimensional representation of the $SO(3,1)$ group.
More concretely, we consider  the tetrad $ e^\mu{}_a(p)$ (or nonholonomic connection) in momentum space, from which the metric can be built
  \begin{equation}
g^{\mu\nu}(p) \,=:\,e^\mu{}_a (p)\eta ^{a b}e^\nu{}_b (p)\,,
\label{eq:metric_tetrad}
\end{equation}
so that the gamma matrices
can be written as
  \begin{equation}\label{eq:gamma_curved}
 \underline{\gamma}^\mu \,:=\,\gamma^a e^\mu{}_a({p})\,.
\end{equation}

As a consequence of the definition \eqref{eq:gamma_curved},
the gamma matrices share several properties with the gamma matrices in curved configuration space~\cite{Birrell:1982ix}. For example, it is immediate to see  that they satisfy a Clifford algebra using the momentum-space metric, i.e.
  \begin{equation}
\lbrace{ \underline{\gamma}^\mu, \underline{\gamma}^\nu\rbrace}_{}\,=\,2 g^{\mu \nu}(p)\mathbb{1}\,,
\end{equation}
where the operator $\lbrace{\cdot,\cdot\rbrace}_{}$ is the anti-commutator. 
Another similarity with the curved configuration space is that, 
if we multiply Eq.~\eqref{eq:Dirac_DSR} by the operator  $\left(  \underline{\gamma}^\nu f_\nu({p})+m\right)$, we reobtain the KG equation derived in the previous section, cf. Eq.~\eqref{eq:KG_DSR}. If instead of considering the Casimir as the squared of the distance in momentum space one chooses any other function of it, this fundamental relationship will not hold. This is for example the case of the Klein--Gordon and Dirac equations obtained in the algebraic approach~\cite{Nowicki:1992if};
a detailed comparison will be performed in  Sec. \ref{sec:dirac_symmetric_basis}.

We close this general discussion by noting that one can derive the modified Dirac  equation from an action principle in momentum space. Indeed, one can define the Dirac adjoint $\bar\psi:=\psi^\dagger \gamma^0$ and the following action,
\begin{align}
 S_{\rm Dirac}:&=\int  {\rm d}^4p\, \sqrt{-g} \bar\psi(-p)\left(  \underline{\gamma}^\mu f_\mu({p})-m\right)\psi(p)\,,
 \label{eq:Dirac_action}
\end{align}
from which the Dirac equation follows when looking for stationary configurations. 

\subsection{Deformed Lorentz invariance and covariance of the Dirac equation}
In the DSR scenario where the metric is given by de Sitter, the Dirac equation in Eq.~\eqref{eq:Dirac_DSR} is invariant under Lorentz transformations.  A proof of this fact can be done  in a way analog to the one  followed in curved configuration space. We will assume that the spinor  $\psi (p)$  transforms linearly (with a matrix $\mathcal{S}$) under a Lorentz transformation $\Lambda$ that takes $p$ to $p^\prime$:
 \begin{align}\label{eq:fermion}
     \psi'(p')\,=\,\mathcal{S}(\Lambda(p))\psi(p)\,.
 \end{align}
 
 By requiring Lorentz invariance, we will then see that $\mathcal{S}$ corresponds to a fermionic (Dirac) representation of the Lorentz group. 
We start by applying $\mathcal{S}$ to Eq.~\eqref{eq:Dirac_DSR}, so that we obtain
   \begin{equation}
 \left(\mathcal{S}  \underline{\gamma}^\mu \mathcal{S}^{-1} f_\mu({p})-m\right)\mathcal{S}\psi(p)
 \,=\, 
 \left(\mathcal{S} \gamma^a \mathcal{S}^{-1}  e^{\prime\rho}{}_b({p}^\prime) f'_\rho ({p}^\prime)-m\right)\psi^\prime (p^\prime)=0\,,
 \label{eq:Dirac_DSR_transformed}
\end{equation}
where we have used appropriate transformations for the vielbein and the functions $f^{\mu}$, 
cf. the definition \eqref{eq:f_definition} (all the quantities in the new system of coordinates are denoted with a prime).

Then, since the proposed diffeomorphism is an isometry of the metric, the vielbein satisfies
\begin{equation}\label{eq:vierbein_isometry}
 e^{\prime\rho}{}_a(p^\prime)\,=\,  e^{\rho}{}_b(p') \Lambda^b{}_a (p^\prime) \,, 
\end{equation}
where $ \Lambda^b{}_a (p^\prime)$ is a (local) Lorentz transformation that may depend on the point $p^\prime $.
Note that this transformation is not the deformed Lorentz transformation obtained from the isometries of the metric.
Considering \eqref{eq:vierbein_isometry} and the symmetry of $f_{\mu}$ (cf. the definition \eqref{eq:f_definition}), we obtain
\begin{equation}
 \Big(\mathcal{S} \gamma^a \mathcal{S}^{-1} \Lambda^b{}_a (p^\prime) e^{\rho}{}_b({p}^\prime)f_\rho ({p}^\prime)-m\Big)\psi^\prime (p^\prime)=0\,.
 \label{eq:Dirac_DSR_transformed2}
\end{equation}

Of course, local Lorentz  transformations near the identity can be expanded in terms of antisymmetric parameters $\epsilon_{ab}:=\eta_{ac} \epsilon^c{}_b$
as customarily,
\begin{equation}
\Lambda^b{}_a (p^\prime) \,=\,\delta^b_a+\epsilon^b{}_a (p^\prime)+\cdots\,. \label{eq:lorentz_tetrad}
\end{equation}
Thus, we can also expand the matrix $\mathcal{S}$ in Eq.~\eqref{eq:Dirac_DSR_transformed} for transformations around the identity; this  allows us to determine the infinitesimal form of $\mathcal{S}$ as a function of the Lorentz coefficients $\epsilon_{ab}(p')$:
   \begin{align}
\mathcal{S}\,&=\, \mathbb{1}-\frac{{\rm i}}{4} \sigma^{ab} \epsilon_{ab}(p^\prime)+\cdots \,,
 \label{eq:transformation_S}
\\
 \sigma^{ab}\,:&=
 \,\frac{{\rm i}}{2}\left[\gamma^a,\gamma^b\right]\,.
 \label{eq:sigma}
\end{align}

More generally, for an arbitrary metric in  momentum space,
 the expression \eqref{eq:Dirac_DSR} is covariant
 once we postulate that $\psi$ is a Dirac fermion transforming according to Eqs. \eqref{eq:fermion} and \eqref{eq:transformation_S} under a local Lorentz transformation.
 This can be immediately proved by noting that,
 because of the properties of the vielbein,  under a joint Lorentz and coordinate transformation we have 
 \begin{align}
  e^{\prime\rho}{}_a({p}^\prime)f^\prime_\rho ({p}^\prime)
 &\,=\,
 (\Lambda^{-1})^{b}{}_{a}(p) e^{\rho}{}_b({p})f_\rho ({p})\,.
 \end{align}
 Replacing this in \eqref{eq:Dirac_DSR} together with \eqref{eq:fermion},
 we arrive to
 \begin{align}
    \Big( \mathcal{S}(\Lambda(p)) \gamma^a \mathcal{S}^{-1}(\Lambda(p)) \Lambda^b{}_a (p) 
    e^{\prime\rho }{}_{b}(p') f^{\prime}_{\rho}(p')-m\Big)\psi^\prime (p^\prime)\,=\,0\,,
 \end{align}
 so that covariance is guaranteed by the usual, albeit local in momentum space compatibility condition for the $\gamma$ matrices and the transformation matrix $\mathcal{S}$:
 \begin{align}
     \mathcal{S}(\Lambda(p)) \gamma^a \mathcal{S}^{-1}(\Lambda(p)) \Lambda^b{}_a (p)\,=\,\gamma^b\,.
 \end{align}

\subsection{Deformed kinematics and choice of tetrad}
Given the important role that the tetrad plays in the Dirac equation, before analyzing some examples we find beneficial to review some results in connection with deformed kinematics. In~\cite{Carmona:2019fwf} one of the present authors has discussed how to obtain a relativistic deformed kinematics  from a maximally de Sitter momentum space. In particular, given a metric, one is able to define different composition laws from different translation generators in momentum space. This is due to the fact that, once the six Lorentz generators $J^{\mu\nu}$ are fixed, there is still an arbitrariness in the definition of the four translation generators $T^\rho$. Indeed, it is always possible to define $T^{\prime \rho}:=T^\rho+c^\rho_{\mu\nu}J^{\mu\nu}$, for arbitrary constants $c^\rho_{\mu\nu}$, so that the $T^{\prime \rho}$ are also generators of isometries in momentum space. Taking this into account,  the most general Lie algebra for the generators of the ten isometries (compatible with rotational symmetry) is given by
\be
[T^0, T^i] \,=\, \frac{c_1}{\Lambda} T^i + \frac{c_2}{\Lambda^2} J^{0i}, \quad\quad\quad [T^i, T^j] \,=\, \frac{c_2}{\Lambda^2} J^{ij}\,.
\label{isoRDK}
\ee
We can distinguish here three kind of kinematics: $\kappa$-Poincaré, for $c_1=\pm 1$ and $c_2=0$;  Snyder~\cite{Battisti:2010sr}, for $c_1=0$ and $c_2=\pm 1$; and hybrid models~\cite{Meljanac:2009ej}, for $c_1\neq 0$ and   $c_2\neq 0$.  

Additionally,  it was noticed in~\cite{Carmona:2019fwf} that the composition law defines unambiguously a tetrad in momentum space. Explicitly,  the isometries associated to quasi-translations can be identified with a deformed composition law $\oplus$, such that 
\be
p\oplus q := T_{p}(q),
\ee
i.e., the composition law is given by a translation with parameters given by the first argument. 
In this scenario it should satisfy  
\be
g_{\mu\nu}\left(p \oplus q\right)\,=\, \frac{\partial \left(p \oplus q\right)_\mu}{\partial q_\rho} g_{\rho \sigma }\left( q\right) \frac{\partial  \left(p \oplus q\right)_\nu}{\partial q_\sigma}\,,
\label{eq:iso}
\ee
so that  taking the limit $q\to 0$ we obtain
\be
g_{\mu\nu}\left(p\right)\,=\,\left. \frac{\partial \left(p \oplus q\right)_\mu}{\partial q_\rho}\right|_{q\to 0} \eta_{\rho \sigma }  \left. \frac{\partial  \left(p \oplus q\right)_\nu}{\partial q_\sigma}\right|_{q\to 0}\,.
\label{eq:iso_tetrad}
\ee
Therefore, once the composition law is known, one can easily associate a tetrad to the momentum space metric, the inverse of the former given by~\cite{Carmona:2019fwf}
\be
{e}_\mu{}^a (p)\,:=\,\delta^a_\nu \left. \frac{\partial \left(p \oplus q\right)_\mu}{\partial q_\nu}\right|_{q\to 0}\,.
\label{eq:tetrad2}
\ee  

One direct consequence is that all the different kinematics which are obtained from the same metric share the deformed dispersion relation (recall that the latter is defined as the square of the distance in momentum space). Instead, these kinematics differentiate one from the other by the corresponding composition law.
As we will see in the next section, considering a tetrad constructed from the  appropriate composition law we will be able to reproduce  results  obtained in the framework of Hopf algebras, 
showing thus the compatibility between both methods.

\subsection{Dirac equation in \texorpdfstring{$\kappa$}{k}-Poincaré kinematics}

\subsubsection{DCL1 basis}
\label{ss:1}
To show the previously-described formalism in action, let us fix the coordinates and consider a particular example of the exact deformed Dirac equation, i.e. keeping all orders in the deformation parameter $\Lambda$. 
In~\cite{Carmona:2019vsh} a really simple basis of $\kappa$-Poincaré kinematics was found, in which the composition law is linear in momenta
\be
\left(p\oplus q\right)_\mu \,=\, p_{\mu} + \left(1 - p_0/\Lambda\right) \,q_{\mu}\,.
\ee 
The momentum metric with these isometries can be defined by the following inverse metric~\cite{Relancio:2021ahm} and inverse tetrad, obtained from~\eqref{eq:tetrad2}:
\begin{align}
g_{\mu \nu}(p)\,:&=\,\eta_{\mu\nu} \left(1-p_0/\Lambda\right)^2\,,\quad p_0<\Lambda\,,
\\
{e}_\mu{}^a(p)\,:&=\,\delta^a_\mu \left(1-p_0/\Lambda\right)\,.
\end{align}
Then, with a direct application of Eq.~\eqref{eq:casimir_metric}, we obtain the Casimir to be\footnote{Upon a redefinition of the mass, the dispersion relation is equivalent to the dispersion relation for time-like deformation of Minkowski space in the right covariant realization of \cite{Juric:2015jxa}.}
\begin{align}
C_{\rm D}^{(simp)}(p)\,=\,\Lambda^2 \arcch^2\left(1+\frac{p^2}{2 \Lambda^2\left(1-p_0/\Lambda\right)}\right)\,,
\label{eq:cas1}
\end{align}
such that the Dirac equation in this basis reads
\be
\label{eq:dirac_simple}
\frac{\Lambda \sqrt{ C_{\text{D}}^{(simp)}(p) }}{\sqrt{\Lambda^2p^2\left(\left(p_0-2\Lambda\right)^2 -\vec{p}^2\right)}} \Big[2 \Lambda\gamma^a \delta_a^{\mu} p_\mu - \gamma^0 (p_0^2+\vec{p}^2)-2 \gamma^i p_i p_0 \Big] \psi(p) \,=\, m\psi(p)\,.
\ee

In spite of the rather complex aspect of this equation, the solutions can be found in a straightforward way. Analogue to what happens in SR, this modified Dirac equation allows four different solutions, namely
\begin{align}
\begin{split}
\psi_1\,=
\begin{pmatrix}
1
\\
0
\\
-\frac{p_3}{p_0+\Lambda \left(e^{m/\Lambda}-1\right)}
\\
-\frac{p_1+\mathi p_2}{p_0+\Lambda \left(e^{m/\Lambda}-1\right)}
\end{pmatrix}
,\,
\psi_2\,=\begin{pmatrix}
0
\\
1
\\
-\frac{p_1-\mathi p_2}{p_0+\Lambda \left(e^{m/\Lambda}-1\right)}
\\
\frac{p_3}{p_0+\Lambda \left(e^{m/\Lambda}-1\right)}
\end{pmatrix}
,\,
\psi_3\,=\begin{pmatrix}
-\frac{p_3}{p_0+\Lambda \left(e^{-m/\Lambda}-1\right)}
\\
-\frac{p_1+\mathi p_2}{p_0+\Lambda \left(e^{-m/\Lambda}-1\right)}
\\
1
\\
0
\end{pmatrix}
,\,
\psi_4\,=\begin{pmatrix}
-\frac{p_1-\mathi p_2}{p_0+\Lambda \left(e^{-m/\Lambda}-1\right)}
\\
\frac{p_3}{p_0+\Lambda \left(e^{-m/\Lambda}-1\right)}
\\
0
\\
1\end{pmatrix}
\end{split}
,\end{align}
where the zeroth-component of the momentum can be obtained from the dispersion relation implied by Eq.~\eqref{eq:cas1}:
\begin{equation}\label{eq:p0_simple}
    p_0\,=\,\Lambda \left(1-\cosh\left(\frac{m}{\Lambda}\right)\left(1{\pm }\sqrt{1-\sech^2\left(\frac{m}{\Lambda}\right)\left(1-\frac{\vec{p}^2}{\Lambda^2}\right)}\right)\right).
\end{equation}
These expressions explicitly display how the deformation occurs in this fermionic case.
Indeed, one can see by expanding the two previous equations in $\Lambda$ that the deformation is proportional to $m/\Lambda$.  This is related to the explicit form of the Casimir in these coordinates, which leads to the  energy-momentum relation of SR for massless particles.
Thus, if we employ the Casimir \eqref{eq:cas1}, hadronic showers \cite{Addazi:2021xuf} (but not gamma ray bursts) may be used to set lower limits on $\Lambda$. This situation could be different for different basis of $\kappa$-Poincaré, i.e. different momentum coordinates.

\subsubsection{Dirac equation in the symmetric basis} \label{sec:dirac_symmetric_basis}

In~\cite{Lukierski:1992dt,Nowicki:1992if} the authors considered the symmetric basis of $\kappa$-Poincaré, where the deformed composition law reads
\be
\left(p \oplus q\right)_0\,=\, p_0+q_0\,,\qquad \left(p \oplus q\right)_i\,=\, p_i e^{q_0/2\Lambda}+q_i e^{-p_0/2\Lambda}\,.
\label{eq:comp_alg}
\ee
Hence, inserting Eq.~\eqref{eq:comp_alg} into Eq.~\eqref{eq:tetrad2}, {the corresponding} tetrad for  momentum space in the symmetric basis of $\kappa$-Poincaré is 
\be
{e}_0{}^0 (p)\,=\,1\,,\qquad {e}_0{}^i (p)\,=\,0\,, \qquad {e}_i{}^0 (p)\,=\,\frac{p_i}{2 \Lambda}\,, \qquad {e}_j{}^i (p)\,=\,\delta^i_j  e^{-p_0/2\Lambda}\,,
\label{eq:tetrad_alg}
\ee
where $i,j = 1,2,3$. From here one can easily obtain the associated metric in momentum space by using the usual relation, given by Eq.~\eqref{eq:metric_tetrad},
\be
g_{00} (p)\,=\,1\,,\qquad g_{0i}(p)=g_{i0} (p)\,=\,\frac{p_i}{2 \Lambda}\,, \qquad g_{ij}(p) \,=\,-\delta^i_j  e^{-p_0/\Lambda}+\frac{p_i p_j}{4 \Lambda^2}\,. 
\label{eq:metric_alg}
\ee

The next step in our construction regards finding the Casimir, defined as the squared of the distance to the origin in momentum space. 
Resorting to Eq.~\eqref{eq:casimir_metric}, it is found to be 
\be
C^{{(S)}}_{\rm D}(p)\,=\, \Lambda^2 \arcch^2 \left(\cosh \left(\frac{p_0}{\Lambda}\right) -\frac{\vec{p}^2}{2 \Lambda^2}\right)\,.
\ee
Finally, the joint use of this Casimir and the tetrad~\eqref{eq:tetrad_alg} in Eq.~\eqref{eq:Dirac_DSR} leads to the following Dirac operator:
\be \label{eq:dirac_symmetric_basis}
\mathcal{D}_{\rm D}^{(S)}\,:=\,\frac{ \sqrt{\frac{C_{\text{D}}^{(S)}(p) }{\Lambda ^2}}}
{2\Lambda  \sinh \left(\sqrt{\frac{C_{\text{D}}^{(S)}(p)}{\Lambda ^2}}\right)}
\left[2 \Lambda  e^{-\frac{p_0}{2 \Lambda }} \gamma^i p_i+\gamma^0  \left(2 \Lambda ^2 \sinh \left(\frac{p_0}{\Lambda }\right)-\vec{p}^2\right)\right]\,.
\ee
Notice that, by construction, the Dirac equation \eqref{eq:dirac_simple} in the simple basis can be mapped to this one by a change of coordinates in momentum space. 

Now we can consider the Dirac equation proposed in~\cite{Nowicki:1992if}, which was obtained by considering the standard real form of the quantum anti-de Sitter algebra, $SO_q(3,2)$. More precisely they have introduced a  finite-dimensional representation of $SO_q(3,2)$ and consistently modified the coproduct. In our language, it should consist in employing the symmetric $\kappa$-Poincaré basis with the Casimir given by expression~\eqref{eq:kg_alg}.
It is important to emphasize that this is not simply a change of coordinates, but a different choice of Casimir.
When inserting Eqs.~\eqref{eq:kg_alg} and~\eqref{eq:tetrad_alg} into \eqref{eq:Dirac_DSR}, we find the Dirac operator
\be
\mathcal{D}^{(S)}_\text{A} \,:=\, \gamma^0 \left(\Lambda \sinh \left(\frac{p_0}{\Lambda}\right)- \frac{\vec{p}^2}{2\Lambda} \right) + e^{-p_0/2\Lambda} p_i \gamma^i\,,
\label{eq:dirac_alg}
\ee
which is indeed the same expression as that obtained in~\cite{Nowicki:1992if}. 
Notice that from the definitions of the Casimirs of Eqs.~\eqref{eq:cas_alg_dis} and~\eqref{eq:kg_alg}, one can see that Eqs.~ \eqref{eq:dirac_symmetric_basis} and \eqref{eq:dirac_alg} are related. Basically,
\be
\mathcal{D}^{(S)}_\text{A} \,=\, \underline{\gamma}^\mu g_{\mu \nu}(p) \frac{\partial C^{(S)}_\text{A}}{\partial p_\nu}\,=\, \underline{\gamma}^\mu g_{\mu \nu}(p) \frac{\partial C^{(S)}_\text{A}}{\partial C^{(S)}_\text{D}}\frac{\partial C^{(S)}_\text{D}}{\partial p_\nu}\,=\, \mathcal{D}^{(S)}_\text{D} \frac{\partial C^{(S)}_\text{A}}{\partial C^{(S)}_\text{D}}\,.
\ee

It is important to note that the ``square'' of the Dirac operator in Eq. \ref{eq:dirac_alg} does not lead to the KG expression \eqref{eq:kg_alg}. 
Indeed, as discussed in~\cite{Nowicki:1992if}, they are related by
\be
\left(
\mathcal{D}^{(S)}_{\text{A}}\right)^2\,=\,C^{(S)}_\text{A}(p)\left(1+\frac{C^{(S)}_\text{A}(p)}{4\Lambda^2}\right)\,.
\ee
Instead,  our proposal \eqref{eq:dirac_symmetric_basis} is such that the KG equation~\eqref{eq:cas_alg_dis} is directly obtained when squaring the Dirac one. 

Another significant fact is that, in obtaining the Dirac equation~\eqref{eq:dirac_alg}, we have employed only the physics in momentum space, without the need of introducing a configuration space. 
If one desires to make contact for example with $\kappa$-Minkowski, one may do that simply by introducing the Fourier transform with the ordering (symmetric, time to the right, time to the left, etc.) of the noncommuting space-time  coordinates appropriate to the chosen coordinates in momentum space~\cite{Mercati:2018hlc}.
However, the Dirac equation~\eqref{eq:dirac_alg} is of more general validity.

At the risk of repeating ourselves, we remark once more that we obtain these results provided we use the particular tetrad given in Eq.~\eqref{eq:tetrad2}. If we considered a different tetrad in momentum space (corresponding to the same metric) we would obtain a different Dirac equation.

\subsection{Relativistic wave equations beyond \texorpdfstring{$\kappa$}{k}-Poincaré kinematics}
The generality of our Dirac equation~\eqref{eq:Dirac_DSR} allows us to consider other kinematics. 
In particular, following the methods described in the preceding sections, 
we will focus on the particular case of Snyder. 
We start by considering the following coordinates for de Sitter
\be 
g_{\mu\nu}(p) \,=\,\eta_{\mu\nu} + p_\mu p_\nu/\Lambda^2\,.
\label{eq:metric_cov_snyder}
\ee
Using these coordinates, it is thus possible to find the dispersion relation as defined in Eq.~\eqref{eq:casimir_metric}~\cite{Pfeifer:2021tas},
\be 
 C_{\text{D}}^{(class)}(p) \,=\, \Lambda^2 \arcsh^2 \left(\frac{\sqrt{p^2}}{\Lambda}\right)\,,
\label{eq:cov_casimir}
\ee
as well as the corresponding composition law associated to Snyder kinematics~\cite{Carmona:2019fwf} in the so-called Maggiore representation~\cite[Eq. (34)]{Battisti:2010sr}
\begin{align}
(p\oplus q)_\mu^{\rm Snyder} \,&=\, p_\mu  \left(\sqrt{1+\frac{q^2}{\Lambda^2}}+\frac{p_{\mu} \eta^{\mu\nu} q_{\nu} }{\Lambda^2\left(1+\sqrt{1+p^2/\Lambda^2}\right)}\right)+ q_\mu \,.
\label{DCLSnyder-1}
\end{align}

Once more, having at disposal the modified composition law, we can employ the recipe given in  Eq.~\eqref{eq:tetrad2}  to straightforwardly obtain the tetrad associated to this kinematics:
\be
{e}_\mu{}^a(p)\,=\, \delta^a_\mu+\frac{p_\mu p_\rho \eta^{\rho\nu}\delta^{a}_\nu}{\Lambda^2 \left(1+\sqrt{1+p^2/\Lambda^2}\right)}\,.
\label{eq:tetrad_snyder}
\ee
Using Eqs.~\eqref{eq:f_definition} and \eqref{eq:Dirac_DSR} one is able to compute the Dirac equation for Snyder kinematics, viz.
\begin{align}
\sqrt{\frac{ C_{\text{D}}^{(class)} }{p^2} } \gamma^a \eta_{ab}\delta^{b}_\nu \eta^{\mu\nu} p_\mu  \psi(p) \,&=\, m\psi(p)\,.
\end{align}

We take now this particular example to stress one our general previous comment: a completely different result is obtained if one considers the composition law of the  $\kappa$-Poincaré kinematics associated to the metric~\eqref{eq:metric_cov_snyder}. It corresponds to the well-known classical basis~\cite{Borowiec2010}, such that
\begin{align}
(p\oplus q)^{\rm \kappa-\text{Poincaré}}_{\mu}\,=\, p_\mu\left(\sqrt{1+\frac{q^2}{\Lambda^2}}+\frac{q_0}{\Lambda}\right)+q_\mu+n_\mu\left[\frac{\sqrt{1+p^2/\Lambda^2}-p_0/\Lambda}{1-\vec{p}^2/\Lambda^2}\left(q_0+\frac{q_\alpha \eta^{\alpha\beta} p_\beta }{\Lambda}\right)-q_0\right] \,,
\end{align}
where $n_{\mu}:=(1,0,0,0)$.
It is immediate to see that the tetrad obtained from~\eqref{eq:tetrad2} is different from the one corresponding to the Snyder model, cf. Eq.~\eqref{eq:tetrad_snyder}. Since the generalized momenta are the same  in both cases\footnote{They are obtained from expression \eqref{eq:f_definition} using  the same dispersion relation~\eqref{eq:cov_casimir}.}, the Dirac equations defined from \eqref{eq:Dirac_DSR} are in its turn different.

\subsection{Discrete symmetries}\label{sec:discrete}
  It is well-known that, in Minkowski space, the Dirac
  equation possesses invariance not only under  continuous Lorentz transformations but also under   discrete ones: parity and time reversal, which connect with the improper and non-orthochronous sectors of the Lorentz group,  and charge conjugation. The corresponding operators are\footnote{Recall that these definitions of time reversal and charge conjugation are valid in Dirac's basis of the gamma matrices.}
   \begin{align}
       \mathcal{P}_0:&\,=\,{\rm i} \gamma^0\,,
       \\
       \mathcal{T}_0:&\,=\,{\rm i} \gamma^1 \gamma^3 \mathcal{K}\,,
       \\
       \mathcal{C}_0:&\,=\, {\rm i} \gamma^2 \mathcal{K}\,,
   \end{align}
   where $\mathcal{K}$ denotes the conjugation operator.
     These operators are intended to act only at the spinor level, while  the coordinates change appropriately when performing a transformation, i.e.
   \begin{align}
       \psi_{\mathcal{A}}(x')\,=\, \mathcal{A} \psi (L_{\mathcal{A}} x)\,,
   \end{align}
   where $L_{\mathcal{A}}$ denotes the corresponding Lorentz transformation acting on the coordinates.
   Notice that, in the flat case, one can recast their action as operators in momentum space. 
   A straightforward computation gives the transformed wave functions
   \begin{align}
        \tilde \psi_{\mathcal{P}}:&\,= \,{\rm i} \gamma^0 \tilde \psi(p_0,-\vec{p})\,, 
    \\
    \tilde \psi_{\mathcal{T}}:&\,= \,{\rm i} \gamma^1\gamma^3 \tilde \psi^*(p_0,-\vec{p})\,,
       \\
       \tilde \psi_{\mathcal{C}}:&\,=\, {\rm i} \gamma^2 \tilde \psi^*(-p)\,,\label{eq:charge_conjugation}
   \end{align}
   from which the action of the relevant operators in momentum space may be deduced. 
In obtaining these transformations, it is crucial that the involved change of coordinates are isometries of the momentum-space Minkowski metric. 
   When going to curved momentum space, we can take these as definitions of the discrete symmetries' operators.
   
   One can thus show  that  $\mathcal{P}$, $\mathcal{T}$  are  discrete symmetries of Dirac's equation if the following condition is met 
   \begin{align}\label{eq:condition_discrete_pt}
        e^{\mu}{}_a(p_0,-\vec{p}) f_{\mu}(p_0,-\vec{p})\,=\,-e^{\mu}{}_a(p) f_{\mu}(p)\,, \quad a\,=\,1,2,3\,.
   \end{align}
 The proof in that case is direct.
  As an example, for parity we can evaluate Eq. \eqref{eq:Dirac_DSR} in $(p_0,-\vec{p})$ and afterwards multiply by $\gamma^0$. Using the anticommuting properties of the gamma matrices one can then show that, if $\tilde \psi$ satisfies Dirac's equation  and condition \eqref{eq:condition_discrete_pt} is satisfied, then  $\tilde\psi_{\mathcal{P}}$ is also a solution of Eq. \eqref{eq:Dirac_DSR}.
  In what concerns the charge conjugation symmetry, since we have not introduced a coupling with electromagnetic fields, we cannot fully tackle the question. However, we can study the behaviour of the free part of Dirac's equation. To prove the invariance under $\mathcal{C}$,  we require
  the assumption
  \begin{align}\label{eq:condition_discrete_c}
       e^{\mu}{}_a(-{p}) f_{\mu}(-{p})\,=\,-e^{\mu}{}_a(p) f_{\mu}(p)\,, \quad a\,=\,0,1,2,3\,.
  \end{align}
  These results are in contrast with the findings in \cite{Andrade:2013oza},
  where the action of $\mathcal{C}$ and $\mathcal{T}$ was found to be incompatible with their modified Dirac equation.
  
 At a first sight, the emergence of  conditions \eqref{eq:condition_discrete_pt} and \eqref{eq:condition_discrete_c} may seem too strong. However, one should recall the analog situation in curved spacetime: not every geometry would admit the definition of such symmetries \cite{Hollands:2002rz},
  even if  a $\mathcal{PCT}$  theorem can be proved in more or less general terms \cite{Hollands:2002rz, Hollands:2009bke}.
In this sense, the situation can be understood as equivalent to the case when an external time-dependent electromagnetic field acting on a particle breaks $\mathcal{T}$ invariance.  

In the same direction, a deeper thought shows that  both conditions \eqref{eq:condition_discrete_pt}-\eqref{eq:condition_discrete_c} are satisfied for rather general tetrads in the $\kappa$-Poincaré scenario (a similar conclusion can be obtained for the Snyder case). We start by noticing that the Casimir is by definition Lorentz invariant and must thus be a function of the rotational invariants $p_0$ and $p^2$ (or equivalently $\vec{p}^2$). 
These are indeed rotational invariants  due to the fact that the time direction is privileged in $\kappa$-Poincaré kinematics, such that while boosts are usually deformed, rotations remain undeformed~\cite{KowalskiGlikman:2002ft}.
Alternatively, as discussed in~\cite{Carmona:2016obd}, one can describe the kinematics of $\kappa$-Poincaré introducing a fixed time-like vector\footnote{
Appearing in tensorial expressions, $n^{\mu}$ can be simply understood as a shorthand to introduce a deformation of usual Lorentz invariance, i.e. a privileged direction in spacetime.
} $n^\mu=(1,0,0,0)$. 
If one wants to avoid the introduction of further elements and also recover the Cartesian SR result in the large $\Lambda$ limit, it is then always possible to recast the gradient of the Casimir ($f^\mu$) as
  \be
f^\mu(p)\,=\,  p^\mu \bar f_1 \left(\frac{p_\alpha n^\alpha}{\Lambda},\frac{p^2}{\Lambda^2}\right)+n^\mu \Lambda  \bar f_2 \left(\frac{p_\alpha n^\alpha}{\Lambda},\frac{p^2}{\Lambda^2}\right)\,,
\label{eq:f_exp}
\ee
  where $\bar f_1$ and $\bar f_2$ are dimensionless functions satisfying the following properties 
    \be
 \bar f_1 \left(0,0\right)\,=\,1\,,\qquad   \lim_{\Lambda\to \infty}\Lambda  \bar f_2 \left(\frac{p_\alpha n^\alpha}{\Lambda},\frac{p^2}{\Lambda^2}\right)\,=\,0\,.
\label{eq:f1f2}
\ee
Notice that, in order to simplify the notation, we are employing the Minkowski metric to raise and lower indices of $p_\mu$ and $n^\mu$.

 We can also write the most general form of the momentum tetrad that respects rotational invariance:
    \begin{align}
    \begin{split}
{e^\mu}_a (p)\,&=\, \delta^\mu_a \bar e_1 \left(\frac{p_\alpha n^\alpha}{\Lambda},\frac{p^2}{\Lambda^2}\right) 
+ \frac{ p^\mu n_a}{\Lambda} \bar  e_2 \left(\frac{p_\alpha n^\alpha}{\Lambda},\frac{p^2}{\Lambda^2}\right) 
+ \frac{n^\mu p_a}{\Lambda} \bar  e_3 \left(\frac{p_\alpha n^\alpha}{\Lambda},\frac{p^2}{\Lambda^2}\right) 
\\
&\hspace{4.5cm}+ \frac{p^\mu p_a}{\Lambda^2} \bar  e_4 \left(\frac{p_\alpha n^\alpha}{\Lambda},\frac{p^2}{\Lambda^2}\right) 
+ n^\mu n_a  \bar  e_5 \left(\frac{p_\alpha n^\alpha}{\Lambda},\frac{p^2}{\Lambda^2}\right)\,.
\label{eq:e_exp}
\end{split}
\end{align}
Since in the limit $\Lambda \to \infty$ we want to recover the canonical tetrad in SR (${e^\mu}_a \to \delta^\mu_a$),
the coefficients $\bar e_i$ will satisfy conditions similar to those in \eqref{eq:f1f2}.
Obviously, relationships between the coefficients $\bar f_i$ and $\bar e_i$ must exist if  Eq.~\eqref{eq:casimir_metric} is to be satisfied.
Analogously, the contraction between the tetrad and $f^\mu$ may be written as 
\be
{e^\mu}_a (p) f_\mu (p)\,=\,p_a \bar a_1 \left(\frac{p_\alpha n^\alpha}{\Lambda},\frac{p^2}{\Lambda^2}\right) +  n_a \Lambda \bar a_2 \left(\frac{p_\alpha n^\alpha}{\Lambda},\frac{p^2}{\Lambda^2}\right)\,,
\label{eq:efproduct}
\ee
  where $\bar a_1$ and $\bar a_2$ are simple  functions of $\bar f_i$ and $\bar e_i$. 
  
  From Eq. \eqref{eq:efproduct} we can directly see that the condition imposed by Eq.~\eqref{eq:condition_discrete_pt} is  satisfied. Additionally, for  Eq.~\eqref{eq:condition_discrete_c}
  to be accomplished, we  need  to replace  $\Lambda \to - \Lambda$ under the action of  $\mathcal{C}$. 
  This can be understood in the following way: taking into account that when applying the symmetry  $\mathcal{C}$ we are changing the sign of the energy, it is quite natural to think that a cutoff scale should change in the same way. Given that such a role is played by the deformation parameter $\Lambda$, we should also do the replacement $\Lambda \to - \Lambda$. 
  In doing so, it is important to mention that, since the momentum scalar of curvature is proportional to $\Lambda^{-2}$, the change $\Lambda \to - \Lambda$ corresponds to an automorphism in the considered space (de Sitter).

  
 {This is not the first time that symmetries involving an action on the deformation parameters have been employed. Indeed, there are several examples in the  quantum groups' literature  proposing that, by introducing a change of sign in the deformation parameter, one can obtain  automorphisms  of the Hopf algebra~\cite{Ballesteros:1993zi,Ballesteros_1994}. This has also been considered in the context of QFT in DSR, as  done in \cite{Kadyshevsky:1983yc} by introducing a pair of scalar fields (linked through the $\Lambda\to -\Lambda$ replacement) and more recently in \cite{Arzano:2020jro}. 
An even more drastic approach was suggested in \cite{Dabrowski:2009mw} where the deformation parameter is promoted to an operator, which is covariant in Wigner's sense.}

  Other recent works in the literature studying discrete symmetries in the DSR context propose a violation of $\mathcal{PCT}$~\cite{Arzano:2019toz,Arzano:2020rzu}. In these works, it was assumed that particles and antiparticles satisfy the same dispersion relation, but the description of the momentum states of the antiparticles is dictated by the antipode of the composition law (as discussed in~\cite{Arzano:2016egk}). This leads to different lifetimes of particles and antiparticles when measured from the laboratory frame of reference, 
  what is readily experimentally testable. 

 In~\cite{Carmona:2021pxw}  a different possible way in which $\mathcal{PCT}$ could be violated  was considered. This violation does not appear in the one-particle system but in the deformed composition law of momenta, affecting in particular in the order in which momenta are composed (the order is important since the composition law is not commutative in $\kappa$-Poincaré kinematics). Therefore, the idea that in the one-particle sector there is no  violation of $\mathcal{PCT}$ is posed.

\section{{Towards a construction of a QFT}}\label{sec:hilbert}

As is well-known, the formal construction and interpretation of the KG theory poses several problems. 
One is able to build a conserved current in SR, $J_{\mu}^{\rm KG}:= {\mathi} \psi^*(x) \overleftrightarrow{\partial}_{\mu} \psi(x)$, where $\overleftrightarrow{\partial}_{\mu}=\overleftarrow{\partial}_{\mu}-\overrightarrow{\partial}_{\mu}$, and even extend its zero component to define a sesquilinear form
\begin{align}\label{eq:scalar_KG}
    (\phi,\psi)_{\rm KG}\,:= \, \mathi \int {\rm d}^3x \big[\partial_t \phi^*(x) \psi(x)-   \phi^*(x) \partial_t\psi(x)\big]\,,
\end{align}
which is conserved in time when $\psi$ and $\phi$ are solutions of the KG equation and sufficiently well-behaved. However, an issue exists, inasmuch as \eqref{eq:scalar_KG} is not positive definite, leading to a failure in its interpretation  as a density and in the definition of a scalar product. 
The way around it, whose generalization enables the construction of QFT in curved spaces, has been to understand that Eq. \eqref{eq:scalar_KG} indeed corresponds to a scalar product if one restricts to positive-energy modes \cite{Wald:1995yp}. This assertion can be immediately seen if one writes a solution $\phi$ of the KG equation in Fourier space;
employing the delta function in \eqref{eq:KG_field} to perform the integral in $p_0$ we obtain
\begin{align}\label{eq:phi_fourier}
    \phi(x)\,=\, \frac{1}{\sqrt{2}(2\pi)^3}\int \measure \left(e^{\mathi (t \freq + \vec{x}\cdot\vec{p}) }\phi_+(\vec{p}) +e^{-\mathi (t \freq+\vec{x}\cdot\vec{p}) }\phi_-(\vec{p})\right)\,,
\end{align}
where $\freq:=\sqrt{\vec{p}^2+m^2}$ is the positive solution one obtains for $p_0$ by enforcing the dispersion relation, while the two  distinct  components $\phi_{\pm}:=\tilde \phi(\pm \freq,\vec{p})$ arise because of  the disconnected support of the delta function in  \eqref{eq:KG_field}. Additionally,  in \eqref{eq:phi_fourier} one can recognize the Lorentz invariant measure.
 Using this expansion one may recast the scalar product as
\begin{align}\label{eq:scalar_KG_Fourier}
    (\phi,\psi)_{\rm  KG}\,=\,\int \measure \big[ \phi_+^*(p) \psi_+(p) -\phi_-^*(p) \psi_-(p)\big]\,,
\end{align}
from which the positive and negative part of the scalar product are evident.

A generalization of Eq. \eqref{eq:phi_fourier} has been developed in~\cite{Arzano:2010jw,Mercati:2018hlc}.
In particular, Ref. \cite{Mercati:2018hlc} includes a term which is proportional to the square root of the product $f^\mu(p) g_{\mu \nu}(p)f^\nu(p)$, which is nothing but our Casimir. Therefore, this term can be easily reabsorbed in a constant.

The fact that an extension of \eqref{eq:scalar_KG_Fourier} may be important also in the construction of a QFT in curved momentum space has been acknowledged in~\cite{Arzano:2010jw}. In order to generalize expression \eqref{eq:scalar_KG_Fourier} in our setup, we need to introduce two changes. First, we rewrite it in the space of four-momenta, including the measure factor and a Dirac delta function that guarantees the satisfaction of Eq. \eqref{eq:KG_DSR}.  A further step is to notice that one can introduce a time-like vector $t_\mu=(1,0,0,0)$. Summing  these elements and given sufficiently well-behaved functions $\phi,\psi\in L^2_{}({\rm dS}_{4})$, i.e.  the space of square-integrable complex-valued functions over ${\rm dS}_4$,  
we can define a positive Hermitian form
\begin{align}\label{eq:scalar_KG_curved}
    (\phi,\psi)_{ C_{\rm D}}\,:=\, 2\int  {\rm d}^4p \,\sqrt{-g} \, \delta\left(C_{\rm D}-m^2 \right) \,  \Theta(f^{\nu} t_{\nu})\, \phi^*(p) \psi(p) \,,
\end{align}
where $\Theta(\cdot)$ is the Heaviside function. It is important to notice that this scalar product is clearly invariant under diffeomorphisms in momentum space. 
An alternative expression can be obtained using the Dirac delta to integrate over $p_0$; one then obtains
\begin{align}\label{eq:scalar_KG_curved_d3p}
    (\phi,\psi)_{ C_{\rm D}}\,=\, \sum_{i}\int  \frac{{\rm d}^3p}{|f^{\nu} t_{\nu}|} \,\sqrt{-g} \,   \Theta(f^{\nu} t_{\nu})\, \phi^*(p) \psi(p)\Big\vert_{p_0=\bar p_0^{(i)}} \,,
\end{align}
where $\bar p_0^{(i)}$ denotes the several solutions to the Casimir equation $C_{\rm D}(\bar p_0^{(i)},\vec{p})-m^2=0$. 
Expression \eqref{eq:scalar_KG_curved_d3p} evidently  resembles the first term in the RHS of \eqref{eq:scalar_KG_Fourier} and
 suggests to proceed in  the following formal way:
one should consider smooth functions with local support on ${\rm dS}_{4}$ and define a projected subspace $\mathscr{S}$ by multiplying by $\Theta(f^{\nu} t_{\nu})$. This is the analogous of choosing the subspace of solutions with positive energy in the standard flat case. Then the Cauchy completion of $\mathscr{S}$ with the scalar product \eqref{eq:scalar_KG_curved} will define a Hilbert space, from which one may try to implement a quantization.

At this point two comments are in order. The description of $\kappa$-Poincaré kinematics customarily involves the introduction of a constant temporal vector\footnote{Strictly speaking, $\kappa$-Poincaré introduces an object which is not a vector under momentum diffeomorphisms. We use the name ``constant temporal vector'' because it is widely employed in this context. }, 
privileging thus the temporal component over the space coordinates, since only the boosts are modified but not the rotations~\cite{KowalskiGlikman:2002ft}.
This means that invariance under diffeomorphisms will be restricted 
to a subset respecting the aforementioned splitting. 
Notice also that the on-shell condition defines a foliation of spacetime with spacelike hyper-surfaces,
whose normal vectors are the generalized momenta $f^\mu$ (which are timelike). This implies that, even if in such case $t_\mu$ would  not transform as a vector under momentum diffeomorphisms, the term $\Theta(f^{\nu} t_{\nu})$ in Eq. \eqref{eq:scalar_KG_curved} would indeed be invariant in the aforementioned restricted sense.

Additionally, as discussed in several papers (see~\cite{Mercati:2018hlc} and references therein), the kinematics of  $\kappa$-Poincaré can be described in just one-half of de Sitter momentum space. Moreover, this half is not closed under the action of Lorentz transformations. 
This restriction is not necessary  in our current discussion, but may be implemented in future discussions involving interactions. 
These ideas are still to be pursued and may offer an alternative to the usual trials of building a QFT in $\kappa$-Poincaré.

   \section{Conclusions}
\label{sec:conclusions}

We have interpreted the relativistic quantum equations, namely the Klein--Gordon and Dirac equations, in a geometric language of curved momentum space, following the ideas of \cite{Carmona:2019fwf, AmelinoCamelia:2011bm,Lobo:2016blj}.
Previous works on the topics were mainly restricted to an algebraic approach. 
In our discussion we have found several features that are worth comment.

First of all, our geometric derivations are shown to be equivalent to the previous algebraic ones \cite{Lukierski:1992dt, Nowicki:1992if}, 
in the sense that we can rederive them with an appropriate choice of the Casimir.
The fact that one can introduce a spinorial representation in two different ways, i.e. by modifying the coproduct of the quantum algebra $SO_q(3,2)$ to include a finite representation or, equivalently, 
by using the tetrad formalism, seems rather striking.
In order to prove this, the correct identification of the functions $f^\mu$ as generalized momenta turned out to be crucial.
Notice that our derivation does not require to make use of a particular choice of configuration space; thus, the Dirac equation \eqref{eq:Dirac_DSR}  applies to the special case of $\kappa$-Minkowski, as well a more general class of theories that may rely on a curved momentum space as a fundamental ingredient.

Second, the selection of the Casimir as the distance in momentum space  establishes a direct connection between the Dirac and Klein--Gordon operators,
being one the ``square'' of the other. 
This is not the case in other trials, where  a nontrivial function  should be introduced to link them  \cite{Nowicki:1992if}.
This is related to the fact that, in principle, one is entitled to choose as Casimir any function of the distance. Our choice can be thus seen as a \emph{minimal choice}, a characteristic which has been already used as guiding principle several times in the past.

Third, we have also discussed the implementation of discrete symmetries in curved momentum space for the fermionic case. We have found that parity, time-reversal and charge conjugation (at the free level) are all preserved in our formalism. This was not the case in previous studies \cite{Andrade:2013oza}, partially because the authors have not realized the need of trading the deformation parameter $\Lambda\to -\Lambda$ under charge conjugation.
Additionally, we have identified the conditions imposed on the vierbeins and the generalized momentum functions, cf. Eqs. \eqref{eq:condition_discrete_pt} and \eqref{eq:condition_discrete_c},
in order for the discrete symmetries to be valid. A deeper interpretation of these expressions and their role in more general geometries is still to be built. In any case, we can affirm that, according to our definition, $\mathcal{PCT}$ is satisfied in this model (at the free level), in accordance with the fundamental role that it plays in standard QFT.

Of course one may try to implement $\mathcal{PCT}$ in other ways. Since we are working with free particles it does not seems natural to include the possible effects of a deformed composition law, which involves dealing with a multi-particle system. With our definition, we are preserving the deformed Lorentz symmetry present in DSR theories: this goes in the same line of the aforementioned Hopf algebraic works~\cite{Ballesteros:1993zi,Ballesteros_1994} and avoids the violation of Lorentz invariance proposed in~\cite{Arzano:2019toz,Arzano:2020rzu}. 

Fourth, we have seen how to apply our proposal to Snyder kinematics,  realizing that the composition law selects a particular tetrad. This shows explicitly that we do not need to constrain ourselves to $\kappa$-Poincaré; instead, we are able to construct the Dirac equation for different kinematical models.

Finally, we have made a first attempt into the identification of the relevant Hilbert space in a quantization process. Further developments in this direction are currently being pursued.

This work admits several generalizations. The most important deals with the incorporation of interactions. 
In this scenario one possibility would be to consider gauge theories, which were recently studied 
both in the quantum $\kappa$-Minkowski (including Becchi--Rouet--Stora--Tyutin symmetries) \cite{Mathieu:2020ccc, Mathieu:2021mxl}
and classical curved-momentum-space level
\cite{Ivetic:2019mkx} (see also \cite{Harikumar:2011um}).
Other possible path regards the consideration of more general metrics in phase space, including a curvature also in spacetime (following the generalization of the momentum metric proposed in~\cite{Relancio:2020zok,Relancio:2020rys,Pfeifer:2021tas}), in order to make contact with recent proposals of (quantum) effective field theories in curved and noncommutative setup \cite{Franchino-Vinas:2019nqy,Franchino-Vinas:2021bcl}.

\section*{Acknowledgments}

 We appreciate useful discussions with Flavio Mercati and valuable comments of Ángel Ballesteros.
SAF is grateful to G. Gori and the Institut für Theoretische Physik, Heidelberg, for their kind hospitality.
SAF acknowledges support from Project 11/X748 and Subsidio a Jóvenes Investigadores 2019, UNLP.  JJR acknowledges support from the
Unión Europea-NextGenerationEU (``Ayudas Margarita Salas para la formación de jóvenes doctores''). This work has been partially supported by Agencia Estatal de Investigaci\'on (Spain)  under grant  PID2019-106802GB-I00/AEI/10.13039/501100011033. 
The authors would like to acknowledge the contribution of the COST Action CA18108 ``Quantum gravity phenomenology in the multi-messenger approach''.

\end{document}